\documentclass[12pt,a4paper]{article}
\pdfoutput=1
\usepackage[format=hang]{caption}
\usepackage{epsfig,amssymb,amsmath,graphicx,subcaption,verbatim,hyperref,xcolor,ulem,epstopdf,psfrag,pstool,braket,array,enumerate,multirow,wrapfig}
\usepackage{graphics,color}

\numberwithin{equation}{section}

\begin{document} 

\newcommand{\be}{\begin{equation}}
\newcommand{\ee}{\end{equation}}
\newcommand{\bea}{\begin{eqnarray}}
\newcommand{\eea}{\end{eqnarray}}
\newcommand{\bean}{\begin{eqnarray*}}
\newcommand{\eean}{\end{eqnarray*}}
\font\upright=cmu10 scaled\magstep1
\font\sans=cmss12
\newcommand{\ssf}{\sans}
\newcommand{\stroke}{\vrule height8pt width0.4pt depth-0.1pt}
\newcommand{\Z}{\hbox{\upright\rlap{\ssf Z}\kern 2.7pt {\ssf Z}}}
\newcommand{\ZZ}{\Z\hskip -10pt \Z_2}
\newcommand{\C}{{\rlap{\upright\rlap{C}\kern 3.8pt\stroke}\phantom{C}}}
\newcommand{\R}{\hbox{\upright\rlap{I}\kern 1.7pt R}}
\newcommand{\HH}{\hbox{\upright\rlap{I}\kern 1.7pt H}}
\newcommand{\CP}{\hbox{\C{\upright\rlap{I}\kern 1.5pt P}}}
\newcommand{\identity}{{\upright\rlap{1}\kern 2.0pt 1}}
\newcommand{\half}{\frac{1}{2}}
\newcommand{\quart}{\frac{1}{4}}
\newcommand{\pr}{\partial}
\newcommand{\bm}{\boldmath}
\newcommand{\I}{{\cal I}} 
\newcommand{\M}{{\cal M}}
\newcommand{\N}{{\cal N}}
\newcommand{\e}{\varepsilon}
\newcommand{\ep}{\epsilon}
\newcommand{\bep}{\mbox{\boldmath $\varepsilon$}}
\newcommand{\Oh}{{\rm O}}
\newcommand{\x}{{\bf x}}
\newcommand{\y}{{\bf y}}
\newcommand{\X}{{\bf X}}
\newcommand{\Y}{{\bf Y}}
\newcommand{\z}{{\bar z}}
\newcommand{\w}{{\bar w}}
\newcommand{\tT}{{\tilde T}}
\newcommand{\tX}{{\tilde\X}}

\thispagestyle{empty}
\rightline{DAMTP-2016-65}
\vskip 3em
\begin{center}
{{\bf \Large Complex Geometry of Nuclei and Atoms
}} 
\\[15mm]

{\bf \large M.~F. Atiyah\footnote{email: m.atiyah@ed.ac.uk}} \\[1pt]
\vskip 1em
{\it 
School of Mathematics, University of Edinburgh, \\
James Clerk Maxwell Building, \\
Peter Guthrie Tait Road, Edinburgh EH9 3FD, U.K.} \\[20pt]

{\bf \large N.~S. Manton\footnote{email: N.S.Manton@damtp.cam.ac.uk}} \\[1pt]
\vskip 1em
{\it 
Department of Applied Mathematics and Theoretical Physics,\\
University of Cambridge, \\
Wilberforce Road, Cambridge CB3 0WA, U.K.}
\vspace{12mm}

%%%%%%%%%%%%%%%%%%%%%%%%%%%%%%%%%%%%%%%%%%%%%%%%%%
\abstract{}
%%%%%%%%%%%%%%%%%%%%%%%%%%%%%%%%%%%%%%%%%%%%%%%%%%
We propose a new geometrical model of matter, in which neutral atoms
are modelled by compact, complex algebraic surfaces. Proton and
neutron numbers are determined by a surface's Chern numbers. 
Equivalently, they are determined by combinations of the 
Hodge numbers, or the Betti numbers. Geometrical 
constraints on algebraic surfaces allow just a finite range of neutron 
numbers for a given proton number. This range encompasses the known isotopes.

\end{center}

\vskip 60pt
Keywords: Atoms, Nuclei, Algebraic Surfaces, 4-Manifolds 
\vskip 5pt

\vfill
\newpage
\setcounter{page}{1}
\renewcommand{\thefootnote}{\arabic{footnote}}

%%%%%%%%%%%%%%%%%%%%%%%%%%%%%%%%%%%%%%%%%%%%%%%%%%%%%%%%%%%%%%%%%%%%%%%%%%%%%
%%%%%%%%%%%%%%%%%%%%%%%%%%%%%%%%%%%%%%%%%%%%%%%%%%%%%%%%%%%%%%%%%%%%%%%%%%%%%

\section{Introduction} 

It is an attractive idea to interpret matter geometrically, and to
identify conserved attributes of matter with topological properties of
the geometry. Kelvin made the pioneering suggestion to model 
atoms as knotted vortices in an ideal fluid \cite{Kel}. Each atom type
would correspond to a distinct knot, and the conservation of atoms in
physical and chemical processes (as understood in the 19th century)
would follow from the inability of knots to change their topology.
Kelvin's model has not survived because atoms are now known to be
structured and divisible, with a nucleus formed of protons and neutrons bound
together, surrounded by electrons. At high energies, these
constituents can be separated. It requires of order 1 eV to remove an
electron from an atom, but a few MeV to remove a proton or
neutron from a nucleus. 

Atomic and nuclear physics has progressed, mainly by treating protons,
neutrons and electrons as point particles, interacting through
electromagnetic and strong nuclear forces \cite{Lil}. Quantum mechanics is an
essential ingredient, and leads to a discrete spectrum of energy
levels, both for the electrons and nuclear particles. The nucleons
(protons and neutrons) are themselves built from three pointlike
quarks, but little understanding of nuclear structure and binding has
so far emerged from quantum chromodynamics (QCD), the theory of quarks.
These point particle models are conceptually not very
satisfactory, because a point is clearly an unphysical idealisation, a
singularity of matter and charge density. An infinite charge density
causes difficulties both in classical electrodynamics \cite{Roh} and in quantum
field theories of the electron. Smoother structures carrying the 
discrete information of proton, neutron and electron number would 
be preferable. 

In this paper, we propose a geometrical model of
neutral atoms where both the proton number $P$ and neutron number $N$ are
topological and none of the constituent particles are pointlike.   
In a neutral atom the electron number is also $P$, because the electron's
electric charge is exactly the opposite of the proton's charge. For
given $P$, atoms (or their nuclei) with different $N$ are known as 
different isotopes. 

A more recent idea than Kelvin's is that of Skyrme, who proposed a 
nonlinear field theory of bosonic pion fields in 3+1 dimensions with 
a single topological invariant, which Skyrme identified with baryon number 
\cite{Sky}. Baryon number (also called atomic mass number) is the sum 
of the proton and neutron numbers, $B = P + N$. Skyrme's baryons are 
solitons in the field theory, so they are
smooth, topologically stable field configurations. Skyrme's model was 
designed to model atomic nuclei, but electrons can be added to produce 
a model of a complete atom. Protons and neutrons can be distinguished 
in the Skyrme model, but only after the internal rotational degrees of 
freedom are quantised \cite{ANW}. This leads to a quantised
``isospin'', with the proton having isospin up ($I_3 = \frac{1}{2}$)
and the neutron having isospin down ($I_3 = -\frac{1}{2}$), where 
$I_3$ the third component of isospin. The model is consistent with 
the well-known Gell-Mann--Nishijima relation \cite{Per}
\be
Q = \frac{1}{2}B + I_3 \,,
\label{chargeisospin}
\ee
where $Q$ is the electric charge of a nucleus (in units of the proton charge) 
and $B$ is the baryon number. $Q$ is integral, because $I_3$ is integer-valued 
(half-integer-valued) when $B$ is even (odd). $Q$ equals the proton number
$P$ of the nucleus and also the electron number of a neutral atom. The neutron
number is $N = \frac{1}{2}B - I_3$. The Skyrme model has had considerable 
success providing models for nuclei \cite{BMSW,LM,Hal,HKM}. Despite
the pion fields being bosonic, the quantised Skyrmions have half-integer 
spin if $B$ is odd \cite{FR}. But a feature of the model is that
proton number and neutron number are not separately topological, and 
electrons have to be added on.

The Skyrme model has a relation to 4-dimensional fields that provides
some motivation for the ideas discussed in this paper. A
Skyrmion can be well approximated by a projection of a 4-dimensional
Yang--Mills field. More precisely, one can take an SU(2) Yang--Mills
instanton and calculate its holonomy along all lines in the
(euclidean) time direction \cite{AM}. The result is a Skyrme field in
3-dimensional space, whose baryon number $B$ equals the instanton number.

So a quasi-geometrical structure in 4-dimensional space (a Yang--Mills 
instanton in flat $\mathbb{R}^4$) can be closely related to
nuclear structure, but still there is just one topological charge. A
next step, first described in the paper \cite{AMS}, was to propose an
identification of smooth, curved 4-manifolds with the fundamental particles in
atoms -- the proton, neutron and electron. Suitable examples of
manifolds were suggested. These manifolds were not all compact, and
the particles they modelled were not all electrically neutral. One of
the more compelling examples was Taub-NUT space as a model for the
electron. By studying the Dirac operator on the Taub-NUT background,
it was shown how the spin of the electron can arise in this context 
\cite{JS1}. There has also been an investigation of multi-electron systems
modelled by multi-Taub-NUT space \cite{FM,Fra}. However, there are some
technical difficulties with the models of the proton and neutron,
and no way has yet been found to geometrically combine protons and 
neutrons into more complicated nuclei surrounded by electrons. Nor is it
clear in this context what exactly should be the topological 
invariants representing proton and neutron number.

A variant of these ideas is a model for the simplest atom, the
neutral hydrogen atom, with one proton and one electron. This appears 
to be well modelled by ${\rm CP}^2$, the complex projective 
plane\footnote{$ \, {\rm CP}^2$ had a different
interpretation in \cite{AMS}.}. The fundamental topological
property of ${\rm CP}^2$ is that it has a generating 2-cycle with
self-intersection 1. The second Betti number is $b_2 = 1$, which 
splits into $b_2^+ = 1$ and $b_2^- = 0$. A complex line in ${\rm CP}^2$ 
represents this cycle, and in the projective plane, two lines always
intersect in one point. A copy of this cycle together with its normal
neighbourhood can be interpreted as the proton part of the atom, whereas the 
neighbourhood of a point dual to this is interpreted as the electron. The 
neighbourhood of a point is just a 4-ball, with a 3-sphere boundary, 
but this is the same as in the Taub-NUT model of the electron, which is
topologically just $\mathbb{R}^4$. The 3-sphere is a twisted circle bundle over
a 2-sphere (the Hopf fibration) and this is sufficient to account for
the electron charge.

In this paper, we have a novel proposal for the proton and neutron
numbers. The 4-manifolds we consider are compact, to model neutral 
atoms. Our previous models always required charged particles to be 
non-compact so that the electric flux could escape to infinity, and 
this is an idea we will retain. We also restrict our manifolds to be 
complex algebraic surfaces, and their Chern numbers will 
be related to the proton and neutron numbers. There 
are more than enough examples to model all currently known isotopes of atoms. 
We will retain ${\rm CP}^2$ as the model for the hydrogen atom.

\section{Topology and Physics of Algebraic Surfaces}

Complex surfaces \cite{BPV} provide a rich supply of compact 4-manifolds. They
are principally classified by two integer topological invariants,
denoted $c_1^2$ and $c_2$. For a surface $X$, $c_1$ and $c_2$ are the 
Chern classes of the complex tangent bundle. $c_2$ is an integer
because $X$ has real dimension 4, whereas the (dual of the) canonical
class $c_1$ is a particular 2-cycle in the second homology group, 
$H_2(X)$. $c_1^2$ is the intersection number of $c_1$ with itself, 
and hence another integer.

\begin{figure}
\centering
\includegraphics[width=5.5in]{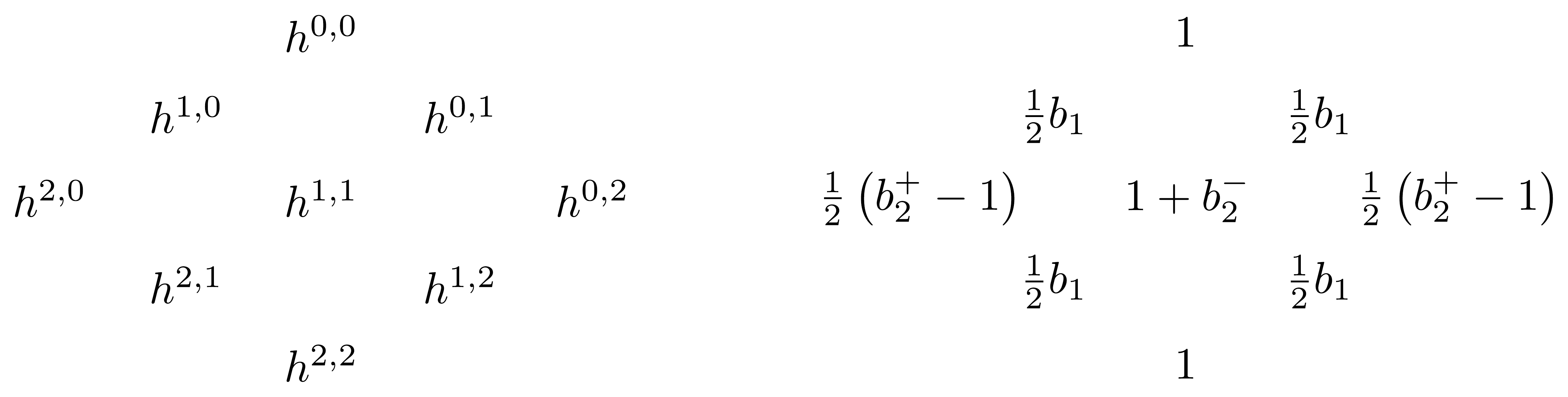}
\vskip 5pt
\caption{The Hodge diamond for a general complex surface (left) and
  its entries in terms of Betti numbers for an algebraic surface (right).}
\label{Hodgediam}
\end{figure}

There are several other topological invariants of a surface
$X$, but many are related to $c_1^2$ and $c_2$. Among the most
fundamental are the Hodge numbers. These are the dimensions of the
Dolbeault cohomology groups of holomorphic forms. In two complex 
dimensions the Hodge numbers are denoted $h^{i,j}$ with 
$0 \le i,j \le 2$. They are arranged in a Hodge diamond, as illustrated in
figure \ref{Hodgediam}. Serre duality, a generalisation of Poincar\'e 
duality, requires this diamond to be unchanged under a $180^\circ$ 
rotation. For a connected surface, $h^{0,0} = h^{2,2} = 1$. 

Complex algebraic surfaces are a fundamental subclass of complex
surfaces \cite{GH,Voi}.
A complex algebraic surface can always be embedded in a complex
projective space ${\rm CP}^n$, and thereby acquires a K\"ahler metric from
the ambient Fubini--Study metric on ${\rm CP}^n$. For any K\"ahler manifold, 
the Hodge numbers have an additional symmetry, $h^{i,j} = h^{j,i}$. For a
surface, this gives just one new relation, $h^{0,1} = h^{1,0}$. Not all
complex surfaces are algebraic: some are still K\"ahler and satisfy
this additional relation, but some are not K\"ahler and do not satisfy
it.

Particularly interesting for us are the holomorphic Euler number
$\chi$, which is an alternating sum of the entries on the top right (or
equivalently, bottom left) diagonal of the Hodge diamond, and the analogous 
quantity for the middle diagonal, which we denote $\theta$. More precisely,
\bea
\chi &=& h^{0,0} - h^{0,1} + h^{0,2} \,, \\
\theta &=& - h^{1,0} + h^{1,1} - h^{1,2} \,.
\eea
(Note the sign choice for $\theta$.) The Euler number $e$ and
signature $\tau$ can be expressed in terms of these as 
\bea
e &=& 2\chi + \theta \,, \\
\tau &=& 2\chi - \theta \,.
\eea
The first of these formulae reduces to the more familiar
alternating sum of Betti numbers $e = b_0 - b_1 + b_2 - b_3 + b_4$,
because each Betti number is the sum of the entries in the
corresponding row of the Hodge diamond. The second formula is the less
trivial Hodge index theorem. $\tau$ is more fundamentally defined by 
the splitting of the second Betti number into positive and negative 
parts, $b_2 = b_2^+ + b_2^-$. Over the reals the intersection form on 
the second homology group $H_2(X)$ is non-degenerate and can be 
diagonalised. $b_2^+$ is then the dimension of the positive subspace, 
and $b_2^-$ the dimension of the negative subspace. The signature is 
$\tau = b_2^+ - b_2^-$. 

The Chern numbers are related to $\chi$ and $\theta$ through the formulae
\be
c_1^2 = 2e + 3\tau = 10\chi - \theta \,, \quad
c_2 = e = 2\chi + \theta \,.
\ee
Their sum gives the Noether formula $\chi =
\frac{1}{12}(c_1^2 + c_2)$, which is always integral.

\begin{figure}
\centering
\includegraphics[width=5.5in]{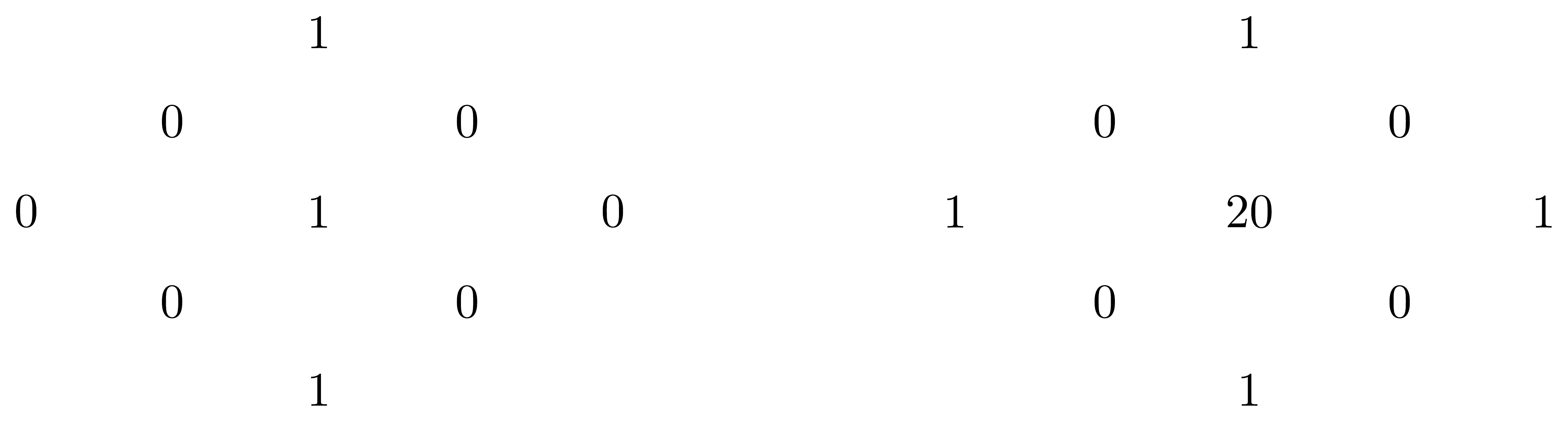}
\vskip 5pt
\caption{Hodge diamonds for the projective plane ${\rm CP}^2$ (left) and for
a ${\rm K3}$ surface (right)}
\label{HodgediamEx}
\end{figure}

For an algebraic surface, there are just three independent Hodge numbers and
they are uniquely determined by the Betti numbers $b_1$, $b_2^+$ and $b_2^-$.
The Hodge diamond must take the form shown on the right in figure 
\ref{Hodgediam}, which gives the correct values for $b_1$, $e$ and 
$\tau$. Note that $b_1$ must be even and $b_2^+$ must be odd. $\chi$
and $\theta$ are now given by
\bea
\chi &=& \frac{1}{2}(1 - b_1 + b_2^+) \,, \\
\theta &=&  1 - b_1 + b_2^- \,.
\eea
If $X$ is simply connected, which accounts for many examples, then 
$b_1 = 0$. Hodge diamonds for the projective plane ${\rm CP}^2$ and for a
${\rm K3}$ surface, both of which are simply connected, are
shown in figure \ref{HodgediamEx}. For the projective plane $\chi = 1$ 
and $\theta = 1$, so $e = 3$ and $\tau = 1$, and for a ${\rm K3}$ surface 
$\chi = 2$ and $\theta = 20$, so $e = 24$ and $\tau = -16$.

Our proposal is to model neutral atoms by complex algebraic surfaces 
and to interpret $\chi$ as proton number $P$, and $\theta$
as baryon number $B$. So neutron number is $N = \theta - \chi$. 
This proposal fits with ${\rm CP}^2$ having $P=1$ and $N=0$. We will see 
later that for each positive value of $P$ there is an interesting, 
finite range of allowed $N$ values. 

In terms of $e$ and $\tau$,
\be
P = \frac{1}{4}(e + \tau) \,, \quad B = \frac{1}{2}(e - \tau) \,, \quad
N = \frac{1}{4}(e - 3\tau) \,.
\ee
Note that for a general, real 4-manifold, these formulae for $P$ and 
$N$ might be fractional, and would need modification.
It is also easy to verify that in terms of $P$ and $N$,
\bea
c_1^2 &=& 9P - N \,, \\
c_2 = e &=& 3P + N \,, \\
\tau &=& P - N \,.
\eea
The simple relation of signature $\tau$ to the difference between 
proton and neutron numbers is striking. If we write 
$N = P + N_{\rm exc} \, $, where $N_{\rm exc}$ denotes the excess of 
neutrons over protons (which is usually zero or positive, but can be 
negative), then $\tau = -N_{\rm exc}$. 

If an algebraic surface $X$ is simply connected then $b_1 = 0$, and in terms 
of $P$ and $N$,
\be
b_2^+ = 2P - 1 \,, \quad b_2^- = P + N - 1 = 2P - 1 + N_{\rm exc} \,.
\label{posnegbetti}
\ee
These formulae will be helpful when we consider 
intersection forms in more detail.

\begin{figure}
\centering
\includegraphics[width=6in]{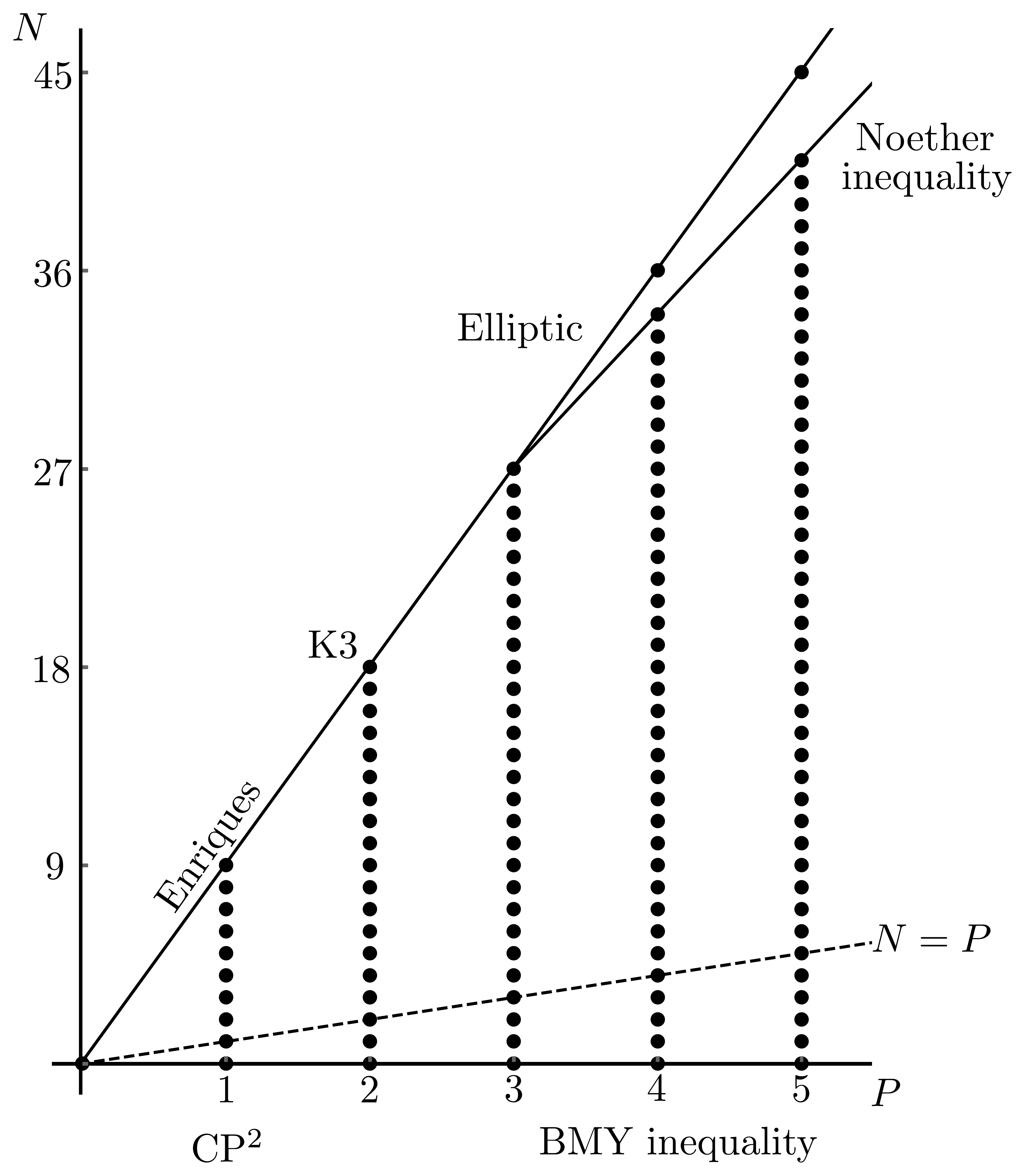}
\vskip 5pt
\caption{Proton numbers $P$, and neutron numbers $N$, for atoms modelled 
as algebraic surfaces. The allowed region is limited by inequalities on 
the Chern numbers, as discussed in the text. Note the change of slope 
from 9 to 7 at the point $P=3$, $N=27$ on the boundary. The line $N=P$
corresponds to surfaces with zero signature, i.e. $\tau = 0$.}
\label{AlgSurfClass}
\end{figure}

The class of surfaces that we will use, as models of atoms, are
those with $c_1^2$ and $c_2$ non-negative. Many of these are
minimal surfaces of general type. Perhaps the most important results 
on the geometry of algebraic surfaces are certain inequalities that the 
Chern numbers of minimal surfaces of general type have to satisfy. The basic
inequalities are that $c_1^2$ and $c_2$ are positive. Also, there is the 
Bogomolov--Miyaoka--Yau (BMY) inequality which requires 
$c_1^2 \le 3c_2$, and finally there is the Noether inequality
$5c_1^2 - c_2 + 36 \ge 0$. These inequalities can be converted into
the following inequalities on $P$ and $N$:
\be
P > 0 \,, \quad 0 \le N < 9P \,, \quad N \le 7P + 6 \,.
\ee
All integer values of $P$ and $N$ satisfying these are
allowed. The allowed region is shown in figure \ref{AlgSurfClass}, 
and corresponds to the allowed region shown on page 229 of \cite{BPV}, 
or in the article \cite{EKWik}. 

There are also the elliptic surfaces (including the Enriques surface
and ${\rm K3}$ 
surface) where $c_1^2 = 0$ and $c_2$ is non-negative, and we shall include 
these among our models. Here, $P \ge 0$ and $N = 9P$, so $c_2 = 12P$ and
$\tau = -8P$. ${\rm CP}^2$ is also allowed, even though it is rational 
and not of general type, because $c_1^2$ and $c_2$ are positive. In
addition to ${\rm CP}^2$, there are further surfaces on the BMY line 
$c_1^2 = 3c_2$ \cite{CS}, which have $P>1$ and $N=0$. 

\begin{figure}
\centering
\includegraphics[width=6in]{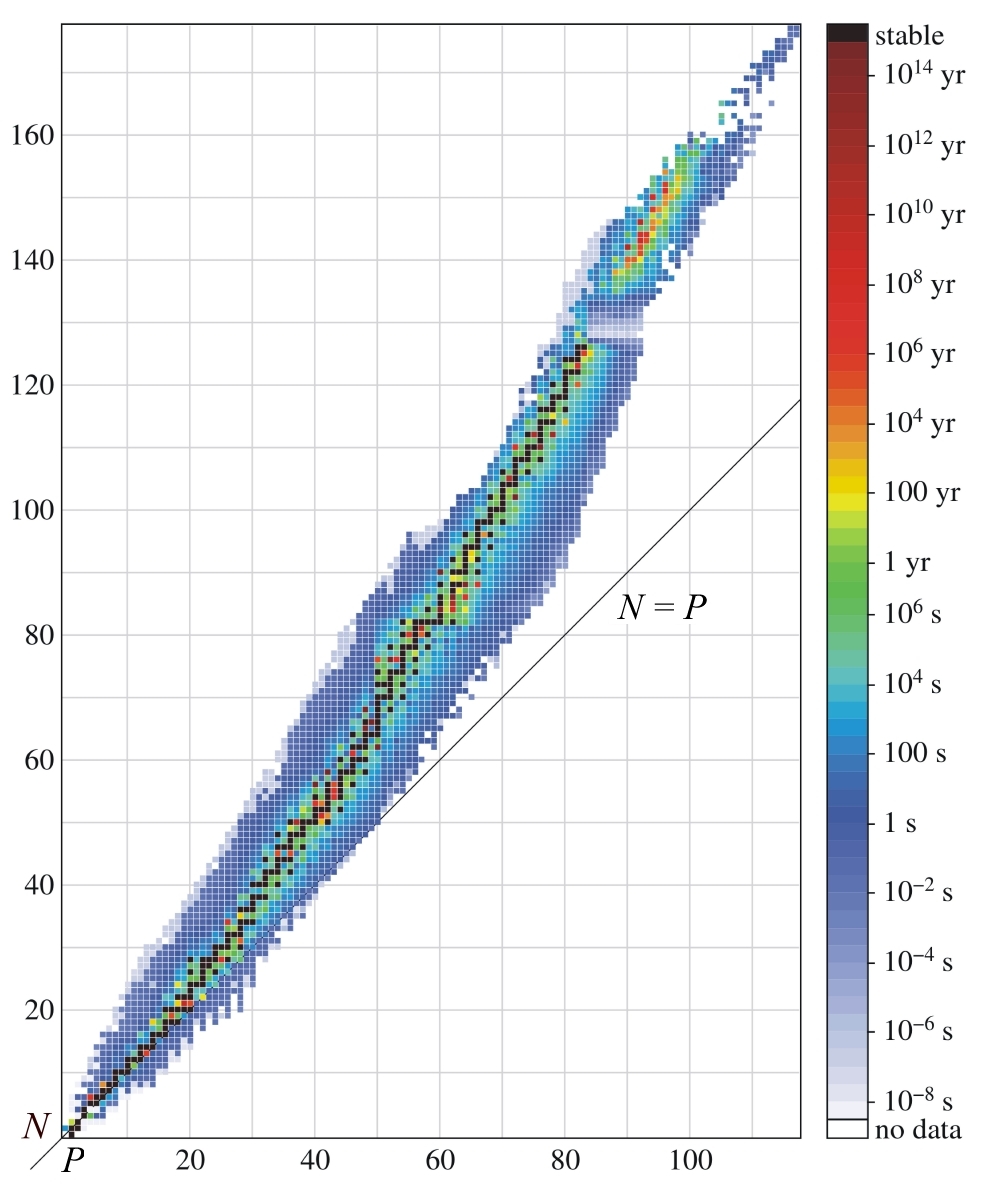}
\vskip 5pt
\caption{Nuclear isotopes. The horizontal axis is proton number $P$
($Z$ in physics notation) and the vertical axis is neutron number 
$N$. The shading (colouring online) indicates the lifetime of each 
isotope, with black denoting stability (infinite lifetime).}
\label{nuclearvalley}
\end{figure}

Physicists usually denote an isotope by proton number
and baryon number, where proton number $P$ is determined 
by the chemical name, and baryon number is $P+N$. 
For example, the notation $^{56}{\rm Fe}$ means
the isotope of iron with $P=26$ and $N=30$. The currently recognised 
isotopes are shown in figure \ref{nuclearvalley}. 

The shape of the allowed region of algebraic surfaces qualitatively matches 
the region of recognised isotopes, and this is the
main justification for our proposal. For example, for $P=1$, the
geometric inequalities allow $N$ to take values from $0$ up to $9$. 
This corresponds to a possible range of hydrogen isotopes from 
$^1{\rm H}$ to $^{10}{\rm H}$. Physically, the
well-known hydrogen isotopes are the proton, deuterium and tritium,
that is, $^1{\rm H}$, $^2{\rm H}$ and $^3{\rm H}$ respectively, but 
nuclear physicists recognise isotopes of a quasi-stable nature 
(resonances) up to $^7{\rm H}$, with $N=6$. 

The minimal models for the common isotopes, the proton alone, and
deuterium, each bound to one electron, are ${\rm CP}^2$ and the complex quadric 
surface ${\rm Q}$. The quadric is the product 
${\rm Q} = {\rm CP}^1 \times {\rm CP}^1$, with $e = 4$
and $\tau = 0$. We shall say more about its intersection form below.

For $P=2$, $N$ is geometrically allowed in the range $0$ to
$18$. The corresponding algebraic surfaces 
should model helium isotopes from $^2{\rm He}$ to $^{20}{\rm He}$. 
Isotopes from $^{3}{\rm He}$ up to $^{10}{\rm He}$ are physically recognised. 
All of these potentially form neutral atoms with two electrons. The 
helium isotope $^{2}{\rm He}$ with no neutrons is not listed in some 
nuclear tables, but there does exist an unbound diproton resonance, 
and diprotons are sometimes emitted when heavier nuclei decay. The 
most common, stable helium isotope is $^4{\rm He}$, with two protons
and two neutrons, but $^3{\rm He}$ is also stable. $^4{\rm He}$ 
nuclei are also called alpha-particles, and play a key role in
nuclear processes and nuclear structure. It is important to have 
a good geometrical model of an alpha-particle, which ideally should match
the cubically symmetric $B=4$ Skyrmion that is a building block for
many larger Skyrmions \cite{BTC,BMS,LM,HKM}.

\section{Valley of Stability}

Running through the nuclear isotopes is the valley of stability \cite{Lil}. In
figure \ref{nuclearvalley}, this is the irregular curved line of 
stable nuclei marked in black. On either side, the 
nuclei are unstable, with lifetimes of
many years near the centre of the valley, reducing to
microseconds further away. Sufficiently far from the centre are the
nuclear drip lines, where a single additional proton or neutron has no
binding at all, and falls off in a time of order $10^{-23}$ seconds.

For small nuclei, for $P$ up to about 20, the valley is centred on the
line $N = P$. In the geometrical model, this line corresponds to surfaces 
with signature $\tau = 0$. For larger $P$, nuclei in the valley have a neutron
excess, $N_{\rm exc}$, which increases slowly from just a few when $P$ is
near 20 to over 50 for the quasi-stable uranium isotopes with $P =
92$, and slightly more for the heaviest artifically produced nuclei
with $P$ approaching 120.

In standard nuclear models, the main effect explaining the valley is
the Pauli principle. Protons and neutrons have a sequence of rather
similar 1-particle states of increasing energy, and just one particle
can be in each state. For given baryon number, the lowest-energy state
has equal proton and neutron numbers, filling the lowest available 
states. If one proton is replaced by one
neutron, the proton state that is emptied has lower energy than the
neutron state that is filled, so the total energy goes up. An
important additional effect is a pairing energy that favours protons
to pair up and neutrons to pair up. Most nuclei with $P$ and $N$ both
odd are unstable as a result. 

For larger values of $P$, the single-particle proton energies tend 
to be higher than the single-particle neutron energies, because 
in addition to the attractive, strong nuclear forces which are
roughly the same for protons and neutrons, there is the electrostatic Coulomb
repulsion that acts between protons alone. This effect becomes
important for nuclei with large $P$, and favours neutron-rich
nuclei. It also explains the instability of
all nuclei with $P$ larger than 83. These nuclei simply split up into
smaller nuclei, either by emitting an alpha-particle, or by fissioning 
into larger fragments. However, the lifetimes
can be billions of years in some cases, which is why uranium,
with $P=92$, is found in nature in relatively large quantities.

Note that if $N = P$, then the electric charge is half the baryon
number, and according to formula (\ref{chargeisospin}), the third component of
isospin is zero. By studying nuclear ground states and excited states,
one can determine the complete isospin, and it is found to be minimal
for stable nuclei. So nuclei with $N = P$ have zero isospin. When the
baryon number is odd, the most stable nuclei have $N$ just one greater
than $P$ (if $P$ is not too large), and the isospin is $\frac{1}{2}$. 
Within the Skyrme model,
isospin arises from the quantisation of internal degrees of freedom,
associated with an SO(3) symmetry acting on the pion
fields. There is an energy contribution proportional to the squared
isospin operator ${\bf I}^2$, analogous to the spin energy proportional to
${\bf J}^2$. In the absence of Coulomb effects, the 
energy is minimised by fixing the isospin to be zero or 
$\frac{1}{2}$. The Coulomb energy competes with isospin, and shifts 
the total energy minimum towards neutron-rich nuclei.

These are the general trends of nuclear energies and
lifetimes. However there is a lot more in the detail. Each isotope has
its own character, depending on its proton and neutron numbers. This is
most clear in the energy spectra of excited states, and the spins of
the ground and excited states. Particularly interesting is the added
stability of nuclei where either the proton or neutron number is magic. 
The smaller magic numbers are 2, 8, 20, 28, 50. It is rather surprising
that protons and neutrons can be treated independently with regard
to the magic properties. This appears to contradict the importance of
isospin, in which protons and neutrons are treated
as strongly influencing each other.       

Particularly stable nuclei are those that are doubly magic, like 
$^4{\rm He}$, $^{16}{\rm O}$, $^{40}{\rm Ca}$ and $^{48}{\rm Ca}$. 
$^{40}{\rm Ca}$ is the largest stable nucleus with $N = P$. 
$^{48}{\rm Ca}$ is also stable, and occurs in small quantities
in nature, but is exceptionally neutron-rich for a relatively small
nucleus.

The important issue for us here is to what extent our proposed geometrical
model based on algebraic surfaces is compatible with these nuclear  
phenomena, not forgetting the electron structure in a neutral
atom. There are some broad similarities. First there is the
``geography'' of surfaces we have discussed above, implying that 
the geometrical inequalities restrict the range of neutron numbers. 
Algebraic geometers also refer to ``botany'', the careful construction 
and study of surfaces with particular topological invariants. The 
patterns are very complicated. Some surfaces are simple to construct, 
others less so, and their internal
structure is very variable. This is analogous to the complications of
the nuclear landscape, and the similar complications (better
understood) of the electron orbitals and atomic shell structure.
	
Rather remarkable is that the line of nuclear stability where $N = P$ 
corresponds to the simple geometrical condition that the signature 
$\tau$ is zero. We have not yet tried to pinpoint an energy function 
on the space of surfaces, but clearly it would be easy to include a dominant
contribution proportional to $\tau^2$, whose minimum would be in the
desired place. Mathematicians have discovered that it is much easier
to construct surfaces on this line, and on the neutron-rich side of
it, where $\tau$ is negative, than on the proton-rich side. There are
always minimal surfaces on the neutron-rich side which are
simply connected, but not everywhere on the proton-rich side. The
geometry of surfaces therefore distinguishes protons from neutrons
rather clearly. This is attractive for the physical interpretation, as
it can be regarded as a prediction of an asymmetry between
the proton and neutron. In standard nuclear physics it is believed that in
an ideal world with no electromagnetic effects, there would be an
exact symmetry between the  proton and neutron, but in reality they
are not the same, partly because of Coulomb energy, but more
fundamentally, because their constituent up (u) and down (d) quarks 
are not identical in their masses, making the proton (uud) less 
massive than the neutron (udd), despite its electric charge.

The geometrical model would need an energy contribution that favours neutrons 
over protons for the larger nuclei and atoms. One possibility has been 
explored by LeBrun \cite{LeB1,LeB2}. This is the infimum, over complex surfaces 
with given topology, of the $L^2$ norm of the scalar curvature. For surfaces
with $b_1$ even, including all surfaces that are simply connected, this 
infimum is simply a constant multiple of $c_1^2$. The scalar curvature 
can be zero for surfaces on the line $c_1^2 = 0$, for example the ${\rm K3}$ 
surface, which is the extreme of neutron-richness, with $P=2$ and 
$N=18$. It would be interesting to consider more carefully 
the energy landscape for an energy that combines $\tau^2$ and a 
positive multiple of the $L^2$ norm of scalar curvature.

\section{Intersection Form}

A complex surface $X$ is automatically oriented, so any pair of 2-cycles has 
an unambiguous intersection number \cite{DK}. Given a basis $\alpha^i$ of
2-cycles for the second homology group $H_2(X)$, the matrix $\Omega^{ij}$ 
of intersection numbers is called the intersection form of $X$. 
$\Omega^{ij} \equiv \Omega(\alpha^i, \alpha^j)$ is the intersection number of 
basis cycles $\alpha^i$ and $\alpha^j$, and the self-intersection number 
$\Omega^{ii}$ is the intersection number of $\alpha^i$ with a generic
smooth deformation of itself. $\Omega$ is a symmetric matrix of
integers, and by Poincar\'e duality it is unimodular (of determinant
$\pm 1$). Over the reals, such a symmetric 
matrix is diagonalisable, and the diagonal entries are either $+1$ or
$-1$. The numbers of each of these are $b_2^+$ and $b_2^-$, respectively, and
we have already given an interpretation of them for simply-connected
algebraic surfaces $X$ in terms of $P$ and $N$ in equation 
(\ref{posnegbetti}) above.

However, diagonalisation over the reals does not make sense for cycles, 
because one can end up with fractional cycles in the new basis. One may only
change the basis of cycles using an invertible matrix of integers,
whose effect is to conjugate $\Omega$ by such a matrix. The
classification of intersection forms is finer over the integers than
the reals. 

For almost all algebraic surfaces, $\Omega$ is indefinite. $b_2^+$ is
always positive, and $b_2^-$ is positive too, except for surfaces with
$b_1 = 0$ and $B = \theta = 1$. So the only surfaces for which the intersection 
form $\Omega$ is definite are ${\rm CP}^2$, and perhaps additionally the fake
projective planes, for which we have not found a physical
interpretation. For ${\rm CP}^2$, with $P=1$ and $N=0$, the 
intersection form is the $1 \times 1$ matrix $\Omega = (1)$. 
Non-degenerate, indefinite forms over the integers have a rather simple
classification. The basic dichotomy is between those that are odd and
those that are even. An odd form is one for which at least one entry 
$\Omega^{ii}$ is odd, or more invariantly, $\Omega(\alpha,\alpha)$ is 
odd for some 2-cycle $\alpha$. An odd form can always be diagonalised, 
with entries $+1$ and $-1$ on the diagonal. 

Even forms are more interesting. Here $\Omega(\alpha,\alpha)$ is even for any
cycle $\alpha$. The simplest example is
\be
\Omega = \begin{pmatrix} 0&1 \\ 1&0 \end{pmatrix} \,.
\label{hypplane}
\ee 
This is the intersection form of the quadric ${\rm Q}$, with the two 
${\rm CP}^1$ factors as basis cycles, $\alpha^1$ and $\alpha^2$. 
If $\alpha = x \alpha^1 + y \alpha^2$ 
then $\Omega(\alpha,\alpha) = 2xy$, so is always
even. Over the reals this form can be diagonalised and has entries $+1$
and $-1$ (the eigenvalues). So it has zero signature. But the 
diagonalisation involves fractional matrices, and is not possible 
over the integers. The intersection form (\ref{hypplane}) is called the
``hyperbolic plane''. A second ingredient in even intersection forms
is the matrix $-E_8$. This is the negative of the Cartan matrix of the
Lie algebra ${\rm E}_8$ (with diagonal entries $-2$). It is even and
unimodular. By itself this form is negative definite, but when
combined with hyperbolic plane components, the result is indefinite,
as needed. The most general (indefinite) even intersection form for an
algebraic surface can be brought to the block diagonal form 
\be
\Omega = l \begin{pmatrix} 0&1 \\ 1&0 \end{pmatrix} \oplus m (-E_8) \,,
\label{evenform}
\ee
with $l>0$ and $m \ge 0$. $l$ must be odd, and the Betti numbers are 
$b_2^+ = l$ and $b_2^- = l + 8m$. The signature is $\tau = -8m$.

For most surfaces, the signature is not a multiple of 8, so
the intersection form is odd. If the signature is a multiple of 8, it
may be even. For given Betti numbers, there could be two distinct
minimal surfaces (or families of these), one with an odd intersection
form, and the other with an even intersection form. We do not know if 
surfaces with both types of intersection form always occur.

We can reexpress these conditions in terms of the physical numbers 
$P$ and $N$. If $N_{\rm exc} = N - P$ is neither zero nor a positive 
multiple of 8, then the intersection form must be odd. If $N = P$, 
then the intersection form can be of the hyperbolic plane type 
$l \begin{pmatrix} 0&1 \\ 1&0 \end{pmatrix}$, with $l = 2P - 1$, 
or it might still be odd. Notice that $l$ is odd, as it must be.
The isotopes for which even intersection forms are possible therefore
include all those with $N = P$. These are numerous. In addition to the
stable isotopes with $N = P$ that occur up to $^{40}{\rm Ca}$, with
$P=20$, there are many that are quasi-stable, like 
$^{52}{\rm Fe}$, with $P=26$. The heaviest recognised isotopes 
with $N = P$ are $^{100}{\rm Sn}$ and perhaps $^{108}{\rm Xe}$, with $P=50$ and
$P=54$. Our geometrical model suggests that the additional stability of these
isotopes is the result of the nontrivial structure of an even
intersection form. 

If $N_{\rm exc} = 8m$ then the intersection form can be of type
(\ref{evenform}), again with $l = 2P - 1$, but it might also be odd. Examples
are the Enriques surface, for which $l=1$ and $m=1$, and the ${\rm K3}$
surface, for which $l=3$ and $m=2$. The potential isotopes corresponding to
these surfaces are $^{10}{\rm H}$ and $^{20}{\rm He}$. These are both
so neutron-rich that they have not been observed, but there are many
heavier nuclei (and corresponding atoms) for which the neutron excess 
$N_{\rm exc}$ is a multiple of 8.

There is some evidence that nuclei whose neutron excess is a multiple
of 8 have additional stability. The most obvious example is $^{48}{\rm Ca}$, 
but this is conventionally attributed to the shell model, as $P=20$
and $N=28$, both magic numbers. A more interesting and less understood 
example is the heaviest known isotope of oxygen, $^{24}{\rm O}$, with 
8 protons and 16 neutrons. This example and others do not obviously 
fit with the shell model. The most stable isotope of iron is 
$^{56}{\rm Fe}$, whose neutron excess is 4, but it is striking that 
$^{60}{\rm Fe}$, whose neutron excess is 8, has a lifetime of over a million
years. Here $P=26$ and $N=34$. $^{64}{\rm Ni}$, also with a neutron excess
of 8, is one of the stable isotopes of nickel. There are also striking
examples of stable or relatively stable isotopes with neutron excesses
of 16 or 24. Some of these are outliers compared to the general trends
in the valley of stability. An example is $^{124}{\rm Sn}$, the
heaviest stable isotope of tin, with $N_{\rm exc} = 24$. A more careful
study would be needed to confirm if the additional stability of
isotopes whose neutron excess is a multiple of 8 is statistically significant.
  
There is no evidence that a neutron deficit of 8 has a stabilising
effect. In fact, almost no nuclei with such a large neutron deficit are
recognised. The only candidate is $^{48}{\rm Ni}$, with the magic
numbers $P=28$ and $N=20$.

\section{Other Surfaces}

In addition to the minimal surfaces of general type there are various
other classes of algebraic surface. Do these have a physical
interpretation?

On a surface $X$ it is usually possible to ``blow up'' one or more
points. The result is not minimal, because a minimal surface, by
definition, is one that cannot be constructed by blowing up points on 
another surface. Blowing up one point increases $c_2$ by 1 and
decreases $c_1^2$ by 1. This is equivalent, in our model, to
increasing $N$ by 1, leaving $P$ unchanged. In other words, one
neutron has been added. Topologically, blowing up is a local process,
equivalent to attaching (by connected sum) a copy of
$\overline{{\rm CP}^2}$. This adds a 2-cycle that has self-intersection
$-1$, but no intersection with any other 2-cycle. The rank
(size) of the intersection form $\Omega$ increases by 1, with an extra $-1$ 
on the diagonal, and the remaining entries of the extra row and 
column all zero. This automatically makes the intersection form odd, 
so any previously even form now becomes diagonalisable.

The physical interpretation seems to be that a neutron has been added,
well separated from any other neutron or proton. This adds a relatively
high energy, more than if the additional neutron were bound into an 
existing nucleus. Minimal algebraic surfaces, and especially those with even
intersection forms, should correspond to tightly bound nuclei
and atoms, having lower energy.

The simplest example is the blow up of one point on ${\rm CP}^2$. The result
is the Hirzebruch surface ${\rm H}_1$, which is a non-trivial 
${\rm CP}^1$ bundle over ${\rm CP}^1$. 
Its intersection form is $\begin{pmatrix} 1&0 \\ 0&-1 \end{pmatrix}$. The
Hirzebruch surface and quadric are both simply connected and have the 
same Betti numbers, $b_2^+ = b_2^- = 1$,
corresponding to $P=1$ and $N=1$, but the intersection form is odd
for the Hirzebruch surface and even for the quadric. The proposed
interpretation is that the Hirzebruch surface represents a separated
proton, neutron and electron, whereas the quadric represents the
deuterium atom, with a bound proton and neutron as its nucleus,
orbited by the electron.

There is an inequality of LeBrun for the $L^2$ norm of the Ricci curvature
supporting this interpretation \cite{LeB1,LeB2}. The norm increases if
points on a minimal surface are blown up, the
increase being a constant multiple of the number of blown-up
points. This indicates that both the norm of the Ricci curvature
and the norm of the scalar curvature, possibly with different coefficients, 
should be ingredients in the physical energy.

So far, we have not considered any surfaces $X$ that could represent 
a single neutron, or a cluster of neutrons. Candidates are the surfaces of Type
VII. These have $c_1^2 = -c_2$, with $c_2$ positive, equivalent to
$P=0$ and arbitrary positive $N$. These surfaces are complex, but are 
not algebraic and do not admit a K\"ahler metric. They are also not 
simply connected. It is important to have a model of a single
neutron. The discussion of blow-ups suggests that $\overline{{\rm CP}^2}$ 
is another possible model. In this case a single neutron would be 
associated with a 2-cycle with self-intersection $-1$, mirroring the 
proton inside ${\rm CP}^2$ being represented by a 2-cycle with 
self-intersection $+1$.

A free neutron is almost stable, having a lifetime of
approximately 10 minutes. There is considerable physical interest in clusters
of neutrons. There is a dineutron resonance similar to the diproton
resonance. Recently there has been some
experimental evidence for a tetraneutron resonance, indicating some
tendency for four neutrons to bind \cite{Kis}. Octaneutron resonances have also
been discussed, but no conclusive evidence for their existence has 
yet emerged. Neutron stars consist of multitudes of neutrons,
accompanied perhaps by a small number of other particles (protons and
electrons), but their stability is only possible because of the
gravitational attraction supplementing the nuclear forces. Standard
Newtonian gravity is of course negligible for atomic nuclei.   

Products of two Riemann surfaces (algebraic curves) of genus 2 or more 
are examples of minimal surfaces of general type, but they are
certainly not simply connected. Their interpretation
as atoms should be investigated. Other surfaces, for example ruled 
surfaces, may have some physical interpretation,
but our formulae would give them negative proton and neutron numbers. 
They do not model antimatter, that is, combinations of antiprotons, 
antineutrons and positrons, because antimatter is probably best
modelled using the complex conjugates of surfaces modelling matter. 
Also bound states of protons and antineutrons, with positive $P$ and
negative $N$, do not seem to exist.

\section{Conclusions}

We have proposed a new geometrical model of matter. It goes beyond our
earlier proposal \cite{AMS} in that it can accommodate far more than
just a limited set of basic particles. In principle, the model can
account for all types of neutral atom. 

Each atom is modelled by a compact, complex algebraic surface, which as a
real manifold is four-dimensional. The physical quantum numbers of proton 
number $P$ (equal to electron number for a neutral atom) and neutron 
number $N$ are expressed in terms of the Chern numbers $c_1^2$ and
$c_2$ of the surface, but they can also be expressed in terms of 
combinations of the Hodge numbers, or of the Betti numbers $b_1$, 
$b_2^+$ and $b_2^-$.

Our formulae for $P$ and $N$ were arrived at by considering the
interpretation of just a few examples of algebraic surfaces -- the
complex projective plane ${\rm CP}^2$, the quadric surface ${\rm Q}$,
and the Hirzebruch surface ${\rm H}_1$. Some consequences, which follow 
from the known constraints on algebraic surfaces, can therefore
be regarded as predictions of the model. Among these are that $P$
is any positive integer, and that $N$ is bounded below by $0$ and bounded
above by the lesser of $9P$ and $7P + 6$. This encompasses all known
isotopes. A most interesting prediction is that
the line $N=P$, which is the centre of the valley of nuclear stability 
for small and medium-sized nuclei, corresponds to the the line $\tau =
0$, where $\tau = b_2^+ - b_2^-$ is the signature. Surfaces with
$\tau$ positive and $\tau$ negative are known to be qualitatively
different, which implies that in our model there is a qualitative difference
between proton-rich and neutron-rich nuclei.

For simply connected surfaces with $b_1 = 0$ (or more generally, if
$b_1$ is held fixed) then an increase of $P$ by 1 corresponds to an
increase of $b_2^+$ by 2. The interpretation is that there are two extra 
2-cycles with positive self-intersection, corresponding to the extra
proton and the extra electron. This matches our earlier models,
where a proton was associated with such a 2-cycle \cite{AMS}, and where
multi-Taub-NUT space with $n$ NUTs modelled $n$ electrons
\cite{FM,Fra}. On the other hand, an increase of $N$ by 1 corresponds
to an increase of $b_2^-$ by 1. This means that a neutron is
associated with a 2-cycle of negative self-intersection, which differs
from our earlier ideas, where a neutron was modelled by a 2-cycle with
zero self-intersection. It appears now that the intersection numbers
are related to isospin (whose third component is $\frac{1}{2}$ for a
proton and $-\frac{1}{2}$ for a neutron) rather than to electric
charge (1 for a proton and 0 for a neutron).

Clearly, much further work is needed to develop these ideas into a
physical model of nuclei and atoms. We have earlier made a few remarks about
possible energy functions for algebraic surfaces. Combinations of the
topological invariants and non-topological curvature integrals should
be explored, and compared with the detailed information on the
energies of nuclei and atoms in their ground states. It will be 
important to account for the quantum mechanical nature of the ground
and excited states, their energies and spins. Discrete energy gaps could
arise from discrete changes in
geometry, for example, by replacing a blown-up surface with a minimal
surface, or by considering the effect of changing $b_1$ while keeping $P$
and $N$ fixed, or by comparing different embeddings of an algebraic
surface in (higher-dimensional) projective space. In some cases there
should be a discrete choice for the intersection form. There are also
possibilities for finding an analogue of a Schr\"odinger equation using
linear operators, like the Laplacian or Dirac operator, acting on
forms or spinors on a surface. Alternatively, the right
approach may be to consider the continuous moduli of surfaces as
dynamical variables, and then quantise these. The moduli should
somehow correspond to the relative positions of the protons, neutrons
and electrons. Some of the ideas just
mentioned have already been investigated in the context of
single particles, modelled by the Taub-NUT space or another non-compact 
4-manifold \cite{JS1,JS2}. Further physical processes, for example,
the fission of larger nuclei, and the binding of atoms into molecules,
also need to be addressed.

Before these investigations can proceed, it will be necessary to
decide what metric structure the surfaces need. Previously, we
generally required manifolds to have a self-dual metric, i.e. to be
gravitational instantons, but this now seems too rigid, as
there are very few compact examples. Requiring a K\"ahler--Einstein 
metric may be more reasonable, although these do not exist for all
algebraic surfaces \cite{Och,Tia}. We plan to pursue the many issues 
raised here in a subsequent paper.

\vspace{7mm}

%%%%%%%%% Acknowledgements %%%%%%%%%%%%%
\section*{Acknowledgements}
%%%%%%%%%%%%%%%%%%%%%%%%%%%%%%%%%%%%%%%%

We are grateful to Chris Halcrow for producing figures 1-3, and Nick
Mee for supplying figure 4.


\begin{thebibliography}{99}

\bibitem{Kel} W. Thomson,
On vortex atoms, 
\textit{Trans. R. Soc. Edin.} {\bf 6} (1867) 94.

\bibitem{Lil} J. Lilley, 
\textit{Nuclear Physics: Principles and Applications},
Chichester: Wiley, 2001.

\bibitem{Roh} F. Rohrlich,
\textit{Classical Charged Particles (3rd ed.)},
Singapore: World Scientific, 2007.

\bibitem{Sky} T.H.R. Skyrme, 
A nonlinear field theory,
\textit{Proc. Roy. Soc. Lond.} {\bf A260} (1961) 127;
A unified field theory of mesons and baryons, 
\textit{Nucl. Phys.} {\bf 31} (1962) 556.

\bibitem{ANW} G.S. Adkins, C.R. Nappi and E. Witten, 
Static properties of nucleons in the Skyrme model,
\textit{Nucl. Phys.} {\bf B228} (1983) 552.

\bibitem{Per} D.H. Perkins, 
\textit{Introduction to High Energy Physics (4th ed.)},
Cambridge: CUP, 2000.

\bibitem{BMSW} R.A. Battye et al.,
Light nuclei of even mass number in the Skyrme model, 
\textit{Phys. Rev.} {\bf C80} (2009) 034323.

\bibitem{LM} P.H.C. Lau and N.S. Manton,
States of Carbon-12 in the Skyrme model,
\textit{Phys. Rev. Lett.} {\bf 113} (2014) 232503.

\bibitem{Hal} C.J. Halcrow,
Vibrational quantisation of the $B=7$ Skyrmion,
\textit{Nucl. Phys.} {\bf B904} (2016) 106. 

\bibitem{HKM} C.J. Halcrow, C. King and N.S. Manton,
A dynamical $\alpha$-cluster model of $^{16}{\rm O}$,
arXiv:1608.05048 (2016).

\bibitem{FR} D. Finkelstein and J. Rubinstein,
Connection between spin, statistics and kinks,
\textit{J. Math. Phys.} {\bf 9} (1968) 1762.

\bibitem{AM} M. Atiyah and N.S. Manton,
Skyrmions from instantons,
\textit{Phys. Lett.} {\bf B222} (1989) 438.

\bibitem{AMS} M. Atiyah, N.S. Manton and B.J. Schroers,
Geometric models of matter,
\textit{Proc. Roy. Soc. Lond.} {\bf A468} (2012) 1252.

\bibitem{JS1} R. Jante and B.J. Schroers,
Dirac operators on the Taub-NUT space, monopoles and SU(2) representations,
\textit{JHEP} 01 (2014) 114.

\bibitem{FM} G. Franchetti and N.S. Manton,
Gravitational instantons as models for charged particle systems,
\textit{JHEP} 03 (2013) 072.

\bibitem{Fra} G. Franchetti,
Harmonic forms on ALF gravitational instantons,
\textit{JHEP} 12 (2014) 075.

\bibitem{BPV} W. Barth, C. Peters and A. Van de Ven,
\textit{Compact Complex Surfaces}, 
Berlin, Heidelberg: Springer, 1984.

\bibitem{GH} P. Griffiths and J. Harris,
\textit{Principles of Algebraic Geometry},
New York, Chichester: Wiley Classics, 1994.

\bibitem{Voi} C. Voisin,
\textit{Hodge Theory and Complex Algebraic Geometry, Vol. I},
Cambridge: CUP, 2002.

\bibitem{EKWik}
\textit{Enriques--Kodaira classification}, 
Wikipedia, 2016.

\bibitem{CS} D.I. Cartwright and T. Steger,
Enumeration of the 50 fake projective planes,
\textit{C. R. Acad. Sci. Paris, Ser. I} {\bf 348} (2010) 11.

\bibitem{BTC} E. Braaten, S. Townsend and L. Carson,
Novel structure of static multisoliton solutions in the Skyrme model,
\textit{Phys. Lett.} {\bf B235} (1990) 147.

\bibitem{BMS} R.A. Battye, N.S. Manton and P.M. Sutcliffe,
Skyrmions and the $\alpha$-particle model of nuclei, 
\textit{Proc. Roy. Soc. Lond.} {\bf A463} (2007) 261.

\bibitem{LeB1} C. LeBrun,
Four-manifolds without Einstein metrics,
\textit{Math. Res. Lett.} {\bf 3} (1996) 133.

\bibitem{LeB2} C. LeBrun,
Ricci curvature, minimal volumes, and Seiberg--Witten theory,
\textit{Invent. Math.} {\bf 145} (2001) 279.

\bibitem{DK} S.K. Donaldson and P.B. Kronheimer, 
\textit{Geometry of Four-Manifolds}, 
Oxford: OUP, 1990.

\bibitem{Kis} K. Kisamori et al.,
Candidate resonant tetraneutron state populated by the 
$^4{\rm He}(^8{\rm He}, \, ^8{\rm Be})$ reaction,
\textit{Phys. Rev. Lett.} {\bf 116} (2016) 052501.

\bibitem{JS2} R. Jante and B.J. Schroers,
Spectral properties of Schwarzschild instantons,
arxiv:1604.06080.

\bibitem{Och} T. Ochiai (ed.),
\textit{K\"ahler Metric and Moduli Spaces: Advanced Studies in Pure
  Mathematics 18-II},
Tokyo: Kinokuniya, 1990.

\bibitem{Tia} G. Tian,
K\"ahler--Einstein metrics with positive scalar curvature,
\textit{Invent. Math.} {\bf 137} (1997) 1.

\end{thebibliography}
\end{document}